\documentclass[prb,twocolumn, superscriptaddress,floatfix]{revtex4}
\usepackage{graphicx}
\usepackage{epstopdf}
\usepackage{bm, amsmath, amssymb}
\usepackage{pdfpages}

\newcommand{\nbar}[0]{\bar{n}}

\newcommand{\x}[0]{\mathrm{X}}
\newcommand{\y}[0]{\mathrm{Y}}
\newcommand{\z}[0]{\mathrm{Z}}

\begin{document}

\title{Observing single quantum trajectories of a superconducting qubit}

\author{K. W. Murch}
\affiliation{Quantum Nanoelectronics Laboratory, Department of Physics, University of California, Berkeley CA 94720}
\author{S. J. Weber}
\affiliation{Quantum Nanoelectronics Laboratory, Department of Physics, University of California, Berkeley CA 94720}
\author{C.  Macklin}
\affiliation{Quantum Nanoelectronics Laboratory, Department of Physics, University of California, Berkeley CA 94720}
\author{I. Siddiqi}
\affiliation{Quantum Nanoelectronics Laboratory, Department of Physics, University of California, Berkeley CA 94720}

\date{\today}

\maketitle

 {\bf  The length of time that a quantum system can exist in a superposition state is determined by how strongly it interacts with its environment. This interaction entangles the quantum state with the inherent fluctuations of the environment. If these fluctuations are not measured, the environment can be viewed as a source of noise, causing random evolution of the quantum system from an initially pure state into a statistical mixture---a process known as decoherence. However, by accurately measuring the environment in real time, the quantum system can be maintained in a pure state and its time evolution described by a \emph{quantum trajectory}\cite{carm93} conditioned on the measurement outcome. We employ weak measurements to monitor a microwave cavity embedding a superconducting qubit and track the individual quantum trajectories of the system. In this architecture, the environment is dominated by the fluctuations of a single electromagnetic mode of the cavity. Using a near-quantum-limited parametric amplifier,\cite{cast08,hatr11para} we selectively measure either the phase or amplitude of the cavity field, and thereby confine trajectories to either the equator or a meridian of the Bloch sphere. We perform quantum state tomography at discrete times along the trajectory to verify that we have faithfully tracked the state of the quantum system as it diffuses on the surface of the Bloch sphere. Our results demonstrate that decoherence can be mitigated by environmental monitoring and validate the foundations of quantum feedback approaches based on Bayesian statistics. \cite{sayr11,vija12,camp13}
  Moreover, our experiments suggest a new route for implementing what Schr\"odinger termed quantum ``steering''\cite{schr35}---harnessing action at a distance to manipulate quantum states via measurement.
 }

If a quantum system and its environment share a common set of stationary states, then a measurement of the environment ultimately leads to projection of the quantum system onto one of its eigenstates. Numerous experiments with photons\cite{guer07}, atoms\cite{kuzm99}, and solid state systems\cite{vija11,hatr13} have elucidated this process, enabling quantum feedback\cite{sayr11,vija12,rist12,camp13} and squeezing. \cite{kuzm00,taka09,schl10,kosc10}   

\begin{figure*}
\includegraphics[angle = 0, width = 1\textwidth]{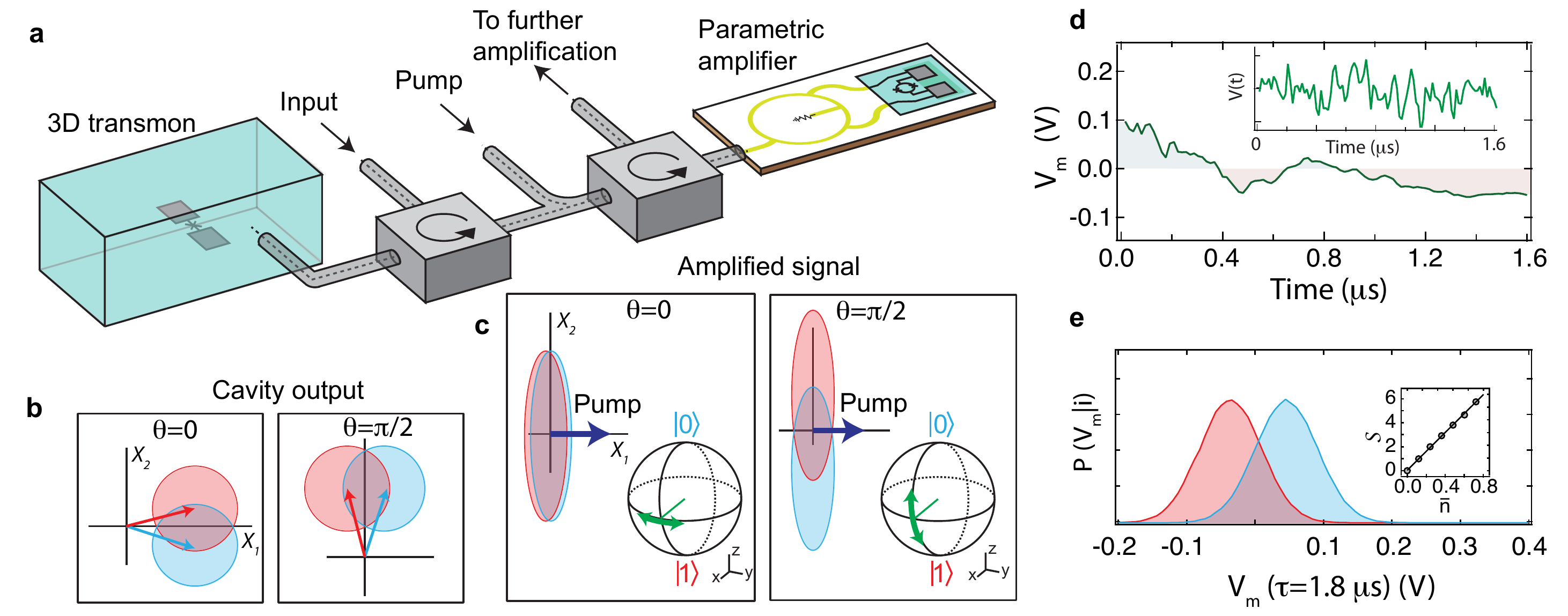}
\caption{\label{fig1} Single quadrature weak measurements.  {\bf a} Our experiment consists of a superconducting transmon qubit dispersivley coupled to a copper waveguide cavity with a coupling rate $\chi/2\pi = -0.49$ MHz.   The cavity port sets the cavity decay rate of  of $\kappa/2\pi = 10.8$ MHz.   Signals that reflect off of the cavity are amplified by a Lumped-element Josephson Parametric Amplifier (LJPA) operating with 10 dB of gain and an instantaneous bandwidth of 20 MHz.    {\bf b} Illustration of the Gaussian variance of measurement signals with phases $\theta = 0$ and $\pi/2$ after reflecting off of the cavity with the qubit in the $|1\rangle$ state (red) and $|0\rangle$ state (blue).  {\bf c} Illustration of the measurement signal after reflecting off of the LJPA.  The amplifier is operated in phase sensitive mode where small signals that are combined in phase(out of phase) with the pump tone are amplified(deamplified) and rotated by $\pi/2$. The backaction of the measurement on a qubit superposition state is indicated on the Bloch sphere.   {\bf d}  A representative integrated measurement signal $V_m(\tau)$ that is obtained when the qubit is prepared in an initial superposition state along the $\hat{x}$ axis of the Bloch sphere.  The inset displays the instantaneous measurement voltage. {\bf e} Histograms of $V_m(\tau=1.8\ \mu$s) for the qubit prepared in the $|1\rangle$ state (red) and $|0\rangle$ state (blue) for a measurement corresponding to $\theta = \pi/2$ with $\mathcal{S} = 3.2$.  The inset displays $\mathcal{S} $ vs.\ $\nbar$, with the solid line indicating the expected dependence for $\eta = 0.49$.
}
\end{figure*}


Our experiment  (Fig.\ 1a) employs a superconducting  transmon qubit\cite{koch07} dispersively coupled to a copper waveguide cavity in the ``3D transmon" architecture.\cite{paik113D} If only the two lowest levels of the transmon are considered, the  qubit-cavity interaction is given by the Hamiltonian $H_\mathrm{int} = -\hbar\chi a^\dagger a \sigma_z$, where $a^\dagger(a)$ is the creation(annihilation) operator for the cavity mode, $\sigma_z$ is the qubit Pauli operator that acts on the qubit state in the energy basis, and $\chi$ is the dispersive coupling rate.  This interaction can be viewed as either a qubit state dependent shift of the cavity frequency of $-\chi\sigma_z$  or a light (or AC Stark) shift of the qubit frequency that depends on the intracavity photon number $\hat{n} = a^\dagger a$.   A microwave tone that probes the cavity near its resonance frequency acquires a qubit state dependent phase shift (Fig.\ 1b).  For $|\chi|\ll\kappa$,  where $\kappa$ is the cavity decay rate, the reflected signal component in quadrature with the input measurement tone contains qubit state information, and the signal component in phase with the measurement tone carries information about the intracavity photon number.  After leaving the cavity, the signal is displaced to the origin of the $X_1 X_2$ plane by a coherent tone and amplified by a near-quantum-limited Lumped-element Josephson Parametric Amplifier (LJPA).\cite{hatr11para}  Phase sensitive operation of the  LJPA permits noiseless amplification of  one quadrature of the reflected measurement signal with corresponding de-amplification of the other quadrature. \cite{cave82,cler10} 

In principle, for an ideal amplifier which adds no noise, the choice of measurement quadrature   determines the type of backaction imparted on a coherent superposition of  qubit states.\cite{koro11}  When the amplified quadrature conveys information about the qubit state, the measurement causes random motions of the qubit  toward its eigenstates,  located at the poles of the Bloch sphere. From the perspective of the qubit, the intracavity photon number \emph{does not fluctuate}.  On the other hand, when the amplified quadrature indicates the intracavity photon number, the phase of the coherent superposition evolves in response to variations of the AC Stark shift of the qubit and superpositions of the measurement eigenstates are not projected.  
We first focus on the case where the amplified quadrature conveys qubit state information which we denote as a  ``$Z$-measurement".   Figure 1d displays a single measurement signal $V_m(\tau)  = \frac{1}{\tau} \int_0^\tau V(t)dt$, where $V(t)$ is the instantaneous measurement voltage, that is obtained when the qubit is initialized in a superposition state $(|0\rangle+|1\rangle)/\sqrt{2}$  along the $\hat{x}$ axis of the Bloch sphere.  As the measurement duration increases, information about the qubit state accumulates. 
The best estimate for the state of the qubit after a weak  measurement can be obtained by Bayes' rule\cite{guer07,koro11,wisebook} which relates  the probability of finding the qubit in state $i$, conditioned on the integrated measurement value $V_m$,
\begin{eqnarray}
P(i | V_m) =\frac{P(i)  P(V_m | i) }{P(V_m) }.
\end{eqnarray}
Probability distributions $P(V_m|i)$ for the integrated measurement value $V_m$ are shown in Fig.\ 1e for the $i=\{|0\rangle,|1\rangle\}$ states.    
$P(i)$ describes the knowledge of the prior distribution and is $1/2$ when the qubit is initialized along $\hat{x}$.  After acquiring a measurement value $V_m$, the state of the system is described by,
\begin{eqnarray}
\z^Z = \tanh(V_m \mathcal{S}/2\Delta V),\  \x^Z = \sqrt{1-(\z^Z)^2} e^{-\gamma \tau}\label{eq:zz}.
\end{eqnarray}
Here we define the expectation values of the Pauli operators conditioned on measurement value $V_m$ as $\x = \langle \sigma_x\rangle |_{V_m}$, $\y = \langle \sigma_y\rangle |_{V_m}$, and $\z = \langle \sigma_z\rangle |_{V_m}$. The superscript $Z$ denotes a $Z$-measurement.  $\mathcal{S} = 64 \tau \chi^2 \nbar \eta/\kappa$ is the dimensionless measurement strength that depends on the measurement duration, $\tau$,  the quantum efficiency of the measurement, $\eta$, and the average intracavity photon number, $\nbar$.  $\mathcal{S}$ can also be related to the  separation of the measurement probability distributions  for the $|0\rangle$ and $|1\rangle$ states ($\Delta V$) and their Gaussian variance ($\sigma^2$), $\mathcal{S} = \Delta V^2/\sigma^2$.     We calibrate $\nbar$ using the measured AC Stark shift of the qubit frequency. From a linear fit of $\mathcal{S}$ vs. $\nbar$ (Fig.\ 1e, inset) we determine $\eta = 0.49$.  For small values of $\mathcal{S}$, an individual measurement does not fully determine the qubit state. When $\mathcal{S}\gg1$, the histograms are well separated and the qubit state can be determined with very high confidence, corresponding to a projective measurement.  

 The exponential decay of coherence in $X^Z$ in equation (\ref{eq:zz})  reflects imperfect knowledge about the state of the environment and leads to qubit dephasing characterized by the rate $\gamma = 8\chi^2\nbar (1-\eta)/\kappa + 1/T_2^*$. The first term in the dephasing rate reflects  measurement induced dephasing\cite{schu05,bois09} originating from the $1-\eta$ of undetected signal.  The second term reflects extra environmental dephasing characterized by $T_2^* = 20\ \mu$s for the qubit.

We now discuss the case of a ``$\phi$-measurement" where the the amplified quadrature is indicative of the fluctuating intracavity photon number.       Each photon that enters the cavity shifts the qubit phase by an average of $4\chi/\kappa$, causing the phase of a coherent superposition of the qubit $|0\rangle$ and $|1\rangle$ states to evolve.  Given $\Delta V$ and $\mathcal{S}$ (obtained from a separate  $Z$-measurement),  $V_m$ can be used to infer the total accrued phase shift.  The evolution of $\x$ and  $\y$ is then given by,
  \begin{eqnarray}
\x^\phi =\cos( \mathcal{S} V_m/(2\Delta V)) e^{-\gamma \tau}\label{eq:xy},\\
\y^\phi =  -\sin(\mathcal{S} V_m/(2\Delta V)) e^{-\gamma \tau} \label{eq:yy}.
\end{eqnarray}
 

   \begin{figure}\begin{center}
\includegraphics[angle = 0, width = 0.5\textwidth]{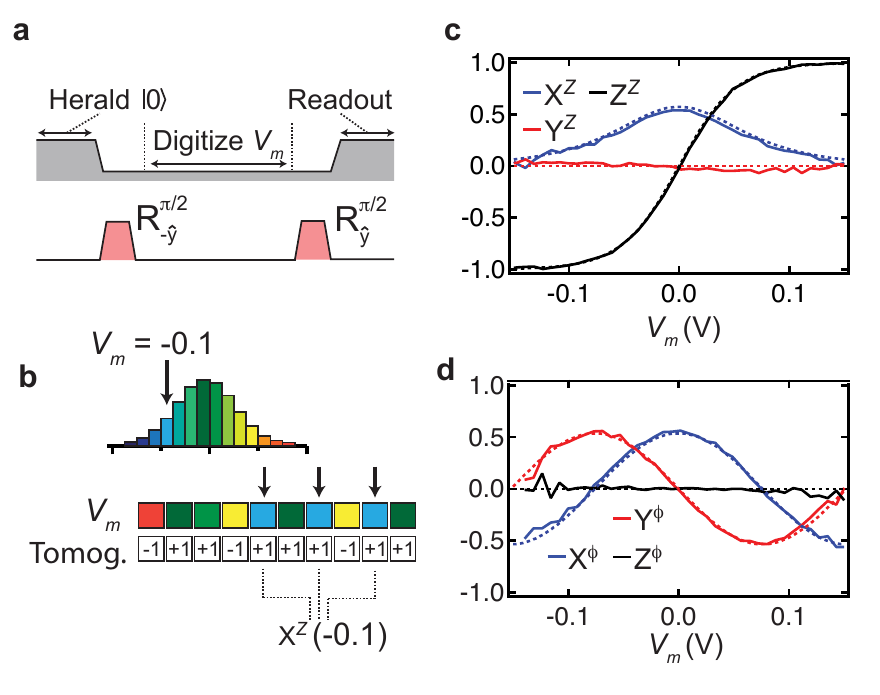}
\end{center}
\caption{\label{fig2} Correlation of tomography results with measurement values.  {\bf a} Experimental sequence for determining $\x^Z$.  An initial strong measurement is used to herald the $|0\rangle$ state followed by a rotation about the $\hat{y}$ axis to prepare the qubit along $\hat{x}$.  A weaker measurement signal is digitized for $1.8\ \mu$s and a final rotation and strong measurement is used to determine the qubit projection along $\hat{x}$.  Similar sequences are used to determine  $\y^Z$ and $\z^Z$. {\bf b}  Tomography correlation procedure.  Different measurement values are indicated as different colors, with the color coded histogram indicating the relative probability of each measurement value. Boxes indicate the measurement value $V_m$ for each experimental repetition and the associated  tomography result is indicated as $\pm 1$.  Tomography results for matching $V_m$ are averaged together to determine $\x^Z$.   {\bf c}  Tomography results vs. $V_m$  for a $Z$-measurement for $\nbar = 0.4$.  The dashed lines are theory curves based on Eq. (\ref{eq:zz}) for $\eta = 0.49$ and $\mathcal{S} =3.15$. {\bf d} Tomography results for a $\phi$-measurement for $\nbar = 0.46$.  The dashed lines are theory curves based on Eq. (\ref{eq:xy}) and (\ref{eq:yy}) for $\eta = 0.49$ and $\mathcal{S} = 3.62$. 
}
\end{figure}

The backaction associated with quadrature specific amplification, as given by equations (2), (3), and (4), is presented in Fig. 2.  To verify these predictions, we conduct an experiment consisting of  three primary actions; we first prepare the qubit along the $\hat{x}$ axis, then we digitize the amplified measurement tone for $1.8\ \mu$s, and finally measure the projection of the qubit state along the $\hat{x},\  \hat{y}$ or $\hat{z}$ axes.  After repeating the experiment sequence $\sim 10^5$ times, we evaluate $\x^{Z,\phi}$, $\y^{Z,\phi}$, and $\z^{Z,\phi}$.  Figure 2c displays the results of this measurement procedure for a $Z$-measurement as a function of $V_m$.    A measurement with  $V_m = 0$ yields no information about the qubit state leaving it unperturbed.  A strongly positive(negative) value of $V_m$, while rare, corresponds to a significant motion of the qubit  toward the $|0\rangle$($|1\rangle$) state.  For the $\phi$-measurement (Fig.\ 2d), $\z^\phi$ is uncorrelated with the measurement signal.  Here $V_m$ conveys information about the phase shift of the qubit state resulting from the fluctuating intracavity photon number.  
For both types of measurement, the results show excellent agreement with theory for $\eta = 0.49$.  





\begin{figure*}
\includegraphics[angle = 0, width = 1\textwidth]{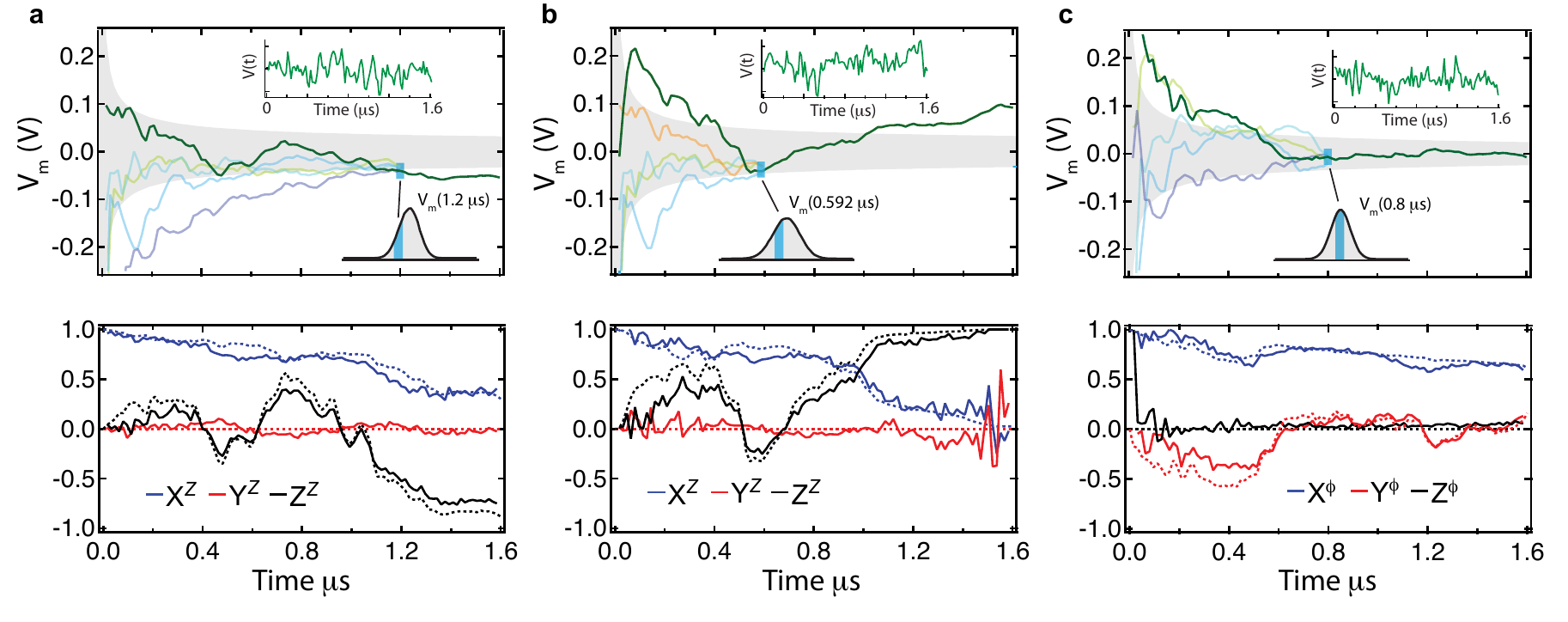}
\caption{\label{fig4} Quantum trajectories.  {\bf a,b}  Individual measurement traces obtained for $Z$-measurements with $\nbar = 0.4$.   The top panel displays $V_m(\tau)$ as a green line, with the inset displaying the instantaneous measurement voltage. The gray region indicates the standard deviation of the distribution of measurement values.  Measurement traces that converge to an integrated value within the blue matching window are used to tomographically reconstruct the trajectory at that time point.  A few different measurement  traces that contribute to the reconstruction at 1.2 $\mu$s ({\bf a}) and 0.592 $\mu$s ({\bf b}) are indicated in pastel colors. The lower insets indicate the distribution of measurement values with the matching window indicated in blue.  Quantum trajectories obtained from analysis of the measurement signal are shown as dashed lines in the lower panel. Solid lines indicate the tomographically reconstructed quantum trajectory based on the ensemble of measurements that are within the matching window of the original measurement signal. {\bf c} Individual measurement traces and associated quantum trajectory obtained for a $\phi$-measurement with $\nbar = 0.4$.}
\end{figure*}

We have so far demonstrated that the integrated measurement signal provides a faithful record of the fluctuations of the environment and the associated motions of the qubit state. Moreover we observe that the direction  of  motion of the qubit state depends on the amplification quadrature. In order to examine quantum trajectories of the system, we divide the measurement signal into successive yet cumulatively integrated segments. As such, the measurement signal can be written as a string $\{V_m(\tau_0), V_m(\tau_1), V_m(\tau_2),\dots\}$, where $\tau_{i+1} - \tau_i = 16$ ns. At each time point,  $V_m(\tau_i)$  can be used to infer a quantum trajectory for the qubit state as it evolves under measurement.  In Fig.\ 3 we present measurement traces along with the quantum trajectory of the system associated with each noisy measurement trace. The trajectories show how the quantum system evolves stochastically from an initial state prepared along $\hat{x}$ toward a final state.  Measurement inefficiency and additional dephasing limits the accuracy with which the state can be tracked, which we display as a gradual shortening of the estimated transverse coherence of the qubit state. 

To verify that we have accurately inferred the quantum trajectory of the system corresponding to a given measurement signal we perform quantum state tomography on an ensemble of experimental iterations with similar measurement values.  
A tomographic reconstruction of the trajectory is obtained by making  measurements of variable duration $\tau_i$ and subsequently measuring the projection of the qubit state along one of the Cartesian axes of the Bloch sphere.   Only measurements with values that are within $\pm \epsilon$ of the target  value $V_m(\tau_i)$ contribute toward determination of the ensemble properties $ \x,\  \y$, and $\z$.  As shown in the upper panels of Fig.\ 3, many different measurement signals that converge to  $V_m(\tau_i)\pm\epsilon$ at $\tau_i$ are used in the tomographic reconstruction. 

Figure 3 a,b display quantum trajectories that are obtained for $Z$-measurements.  The reconstructed trajectories based on ensemble measurements, shown as solid lines, are in reasonable agreement with the quantum trajectories determined from a single measurement record and reproduce many of the minute motions of the qubit as it ultimately evolves toward its eigenstates of measurement.  Some trajectories highlight the concept of quantum measurement reversal:\cite{koro06,katz06,katz08,kim11} in Fig.\ 3a, after $\sim 400,\  600,$ and $1000$ ns of measurement the qubit state has been returned nearly to its original state, effectively ``reversing" the preceding partial collapse of the qubit wavefunction.  
 In Fig.\ 3c we display the measurement record we obtain from a $\phi$-measurement.  The resulting quantum trajectory is confined to motions along the equator of the Bloch sphere.

Full control over the environment of a quantum system allows for the mitigation of decoherence through accurate monitoring of fluctuations of the environment. 
Realizing potential applications of quantum feedback \cite{sayr11,vija12}  in quantum metrology and information science  will require measurement efficiencies approaching unity. While measurement schemes based on projective measurements on ancilla qubits obtain measurement efficiencies\cite{groe13} $>0.9$, the measurement efficiency $\eta = 0.49$ presented here is among the highest reported values for a continuous variable.\cite{vija12, sayr11,hatr13,camp13} As we discuss in Supplemental Information, further  improvements in this efficiency will enable us steer quantum systems through measurement.\cite{wise12}

{\bf Methods Summary}

The qubit consists of two aluminum paddles connected by a double-angle-evaporated aluminum SQUID deposited on double-side-polished silicon, and is characterized by charging and Josephson energies $E_\mathrm{c}/h =200 $ MHz and $E_\textrm{J}/h = 11$ GHz respectively.  The qubit is operated with negligible flux threading the SQUID loop with $\omega_\mathrm{q}/2\pi =3.999$ GHz. The qubit is located off center of a $6.8316$ GHz copper waveguide cavity. 

The LJPA consists of a two junction SQUID, formed from $2 \ \mu$A Josephson junctions, shunted by $3$ pF of capacitance, and is flux biased to provide $10$ dB of gain at the cavity resonance frequency.  The LJPA is pumped by two sidebands that are equally spaced 300 MHz above and below the cavity resonance.  A second following LJPA provides additional gain. A detailed experimental schematic is shown in supplemental Fig.\ S1.

Experiment sequences start with an 800 ns readout with $\mathcal{S} = 42 $ that is used to herald the $|0\rangle$ state at the beginning of the experiment.  A sample herald histogram is shown in supplemental Fig. S2. Because $|\chi|\ll\kappa$, several peaks are visible corresponding to the many energy levels of the transmon qubit.   After preparing the $|0\rangle$ state we perform a 16 ns $\pi/2$ rotation about the $-\hat{y}$ axis to initialize the qubit along the $\hat{x}$ axis.  Following a variable duration we perform quantum state tomography, which consists of either rotations about the $\hat{x}$ axis, $\hat{y}$ axis or no rotation and a second 800 ns readout with $\mathcal{S} =42 $.  A fraction ($\sim4\%$) of the final readouts find the qubit outside of the $\{|0\rangle,|1\rangle\}$ manifold and were disregarded in the analysis.  Tomography results are corrected for the readout fidelity of $95\%$.

{\bf Acknowledgments}

We thank H. Wiseman, A. N. Korotkov, E. M. Levenson-Falk, and N. Roch for useful discussions and R. Vijay for contributions to preliminary experiments.  This research was supported in part by the Office of Naval Research (ONR) and the Office of the Director of National Intelligence (ODNI), Intelligence Advanced Research Projects Activity (IARPA), through the Army Research Office. All statements of fact, opinion or conclusions contained herein are those of the authors and should not be construed as representing the official views or policies of IARPA, the ODNI, or the US Government. 






\begin{thebibliography}{10}
\expandafter\ifx\csname url\endcsname\relax
  \def\url#1{\texttt{#1}}\fi
\expandafter\ifx\csname urlprefix\endcsname\relax\def\urlprefix{URL }\fi
\providecommand{\bibinfo}[2]{#2}
\providecommand{\eprint}[2][]{\url{#2}}

\bibitem{carm93}
\bibinfo{author}{Carmichael, H.}
\newblock \emph{\bibinfo{title}{An Open Systems Approach to Quantum Optics}}
  (\bibinfo{publisher}{Springer-Verlag}, \bibinfo{year}{1993}).

\bibitem{cast08}
\bibinfo{author}{Castellanos-Beltran, M.~A.}, \bibinfo{author}{Irwin, K.~D.},
  \bibinfo{author}{Hilton, G.~C.}, \bibinfo{author}{Vale, L.~R.} \&
  \bibinfo{author}{Lehnert, K.~W.}
\newblock \bibinfo{title}{Amplification and squeezing of quantum noise with a
  tunable Josephson metamaterial}.
\newblock \emph{\bibinfo{journal}{Nature Physics}}
  \textbf{\bibinfo{volume}{4}}, \bibinfo{pages}{929--931}
  (\bibinfo{year}{2008}).

\bibitem{hatr11para}
\bibinfo{author}{Hatridge, M.}, \bibinfo{author}{Vijay, R.},
  \bibinfo{author}{Slichter, D.~H.}, \bibinfo{author}{Clarke, J.} \&
  \bibinfo{author}{Siddiqi, I.}
\newblock \bibinfo{title}{Dispersive magnetometry with a quantum limited SQUID
  parametric amplifier}.
\newblock \emph{\bibinfo{journal}{Phys. Rev. B}} \textbf{\bibinfo{volume}{83}},
  \bibinfo{pages}{134501} (\bibinfo{year}{2011}).

\bibitem{sayr11}
\bibinfo{author}{Sayrin, C.} \emph{et~al.}
\newblock \bibinfo{title}{Real-time quantum feedback prepares and stabilizes
  photon number states}.
\newblock \emph{\bibinfo{journal}{Nature}} \textbf{\bibinfo{volume}{477}},
  \bibinfo{pages}{73} (\bibinfo{year}{2011}).

\bibitem{vija12}
\bibinfo{author}{Vijay, R.} \emph{et~al.}
\newblock \bibinfo{title}{Stabilizing rabi oscillations in a superconducting
  qubit using quantum feedback}.
\newblock \emph{\bibinfo{journal}{Nature}} \textbf{\bibinfo{volume}{490}},
  \bibinfo{pages}{77} (\bibinfo{year}{2012}).

\bibitem{camp13}
\bibinfo{author}{Campagne-Ibarcq, P.} \emph{et~al.}
\newblock \bibinfo{title}{Stabilizing the trajectory of a superconducting qubit
  by projective measurement feedback}.
\newblock \emph{\bibinfo{journal}{arXiv:1301.6095}}  (\bibinfo{year}{2013}).

\bibitem{schr35}
\bibinfo{author}{Schr\"odinger, E.}
\newblock \bibinfo{title}{Die gegenw\"artige situation in der quantenmechanik}.
\newblock \emph{\bibinfo{journal}{Naturwissenschaften}}
  \textbf{\bibinfo{volume}{23}}, \bibinfo{pages}{807--812,823--824,844--849}
  (\bibinfo{year}{1935}).

\bibitem{guer07}
\bibinfo{author}{Guerlin, C.} \emph{et~al.}
\newblock \bibinfo{title}{Progressive field-state collapse and quantum
  non-demolition photon counting}.
\newblock \emph{\bibinfo{journal}{Nature}} \textbf{\bibinfo{volume}{448}},
  \bibinfo{pages}{889} (\bibinfo{year}{2007}).

\bibitem{kuzm99}
\bibinfo{author}{Kuzmich, A.} \emph{et~al.}
\newblock \bibinfo{title}{Quantum nondemolition measurements of collective
  atomic spin}.
\newblock \emph{\bibinfo{journal}{Phys. Rev. A}} \textbf{\bibinfo{volume}{60}},
  \bibinfo{pages}{2346--2350} (\bibinfo{year}{1999}).

\bibitem{vija11}
\bibinfo{author}{Vijay, R.}, \bibinfo{author}{Slichter, D.~H.} \&
  \bibinfo{author}{Siddiqi, I.}
\newblock \bibinfo{title}{Observation of quantum jumps in a superconducting
  artificial atom}.
\newblock \emph{\bibinfo{journal}{Phys. Rev. Lett.}}
  \textbf{\bibinfo{volume}{106}}, \bibinfo{pages}{110502}
  (\bibinfo{year}{2011}).

\bibitem{hatr13}
\bibinfo{author}{Hatridge, M.} \emph{et~al.}
\newblock \bibinfo{title}{Quantum back-action of an individual
  variable-strength measurement}.
\newblock \emph{\bibinfo{journal}{Science}} \textbf{\bibinfo{volume}{339}},
  \bibinfo{pages}{178--181} (\bibinfo{year}{2013}).

\bibitem{rist12}
\bibinfo{author}{Rist\`e, D.}, \bibinfo{author}{Bultink, C.~C.},
  \bibinfo{author}{Lehnert, K.~W.} \& \bibinfo{author}{DiCarlo, L.}
\newblock \bibinfo{title}{Feedback control of a solid-state qubit using
  high-fidelity projective measurement}.
\newblock \emph{\bibinfo{journal}{Phys. Rev. Lett.}}
  \textbf{\bibinfo{volume}{109}}, \bibinfo{pages}{240502}
  (\bibinfo{year}{2012}).

\bibitem{kuzm00}
\bibinfo{author}{Kuzmich, A.}, \bibinfo{author}{Mandel, L.} \&
  \bibinfo{author}{Bigelow, N.~P.}
\newblock \bibinfo{title}{Generation of spin squeezing via continuous quantum
  nondemolition measurement}.
\newblock \emph{\bibinfo{journal}{Phys. Rev. Lett.}}
  \textbf{\bibinfo{volume}{85}}, \bibinfo{pages}{1594--1597}
  (\bibinfo{year}{2000}).

\bibitem{taka09}
\bibinfo{author}{Takano, T.}, \bibinfo{author}{Fuyama, M.},
  \bibinfo{author}{Namiki, R.} \& \bibinfo{author}{Takahashi, Y.}
\newblock \bibinfo{title}{Spin squeezing of a cold atomic ensemble with the
  nuclear spin of one-half}.
\newblock \emph{\bibinfo{journal}{Phys. Rev. Lett.}}
  \textbf{\bibinfo{volume}{102}}, \bibinfo{pages}{033601}
  (\bibinfo{year}{2009}).

\bibitem{schl10}
\bibinfo{author}{Schleier-Smith, M.~H.}, \bibinfo{author}{Leroux, I.~D.} \&
  \bibinfo{author}{Vuleti\ifmmode~\acute{c}\else \'{c}\fi{}, V.}
\newblock \bibinfo{title}{States of an ensemble of two-level atoms with reduced
  quantum uncertainty}.
\newblock \emph{\bibinfo{journal}{Phys. Rev. Lett.}}
  \textbf{\bibinfo{volume}{104}}, \bibinfo{pages}{073604}
  (\bibinfo{year}{2010}).

\bibitem{kosc10}
\bibinfo{author}{Koschorreck, M.}, \bibinfo{author}{Napolitano, M.},
  \bibinfo{author}{Dubost, B.} \& \bibinfo{author}{Mitchell, M.~W.}
\newblock \bibinfo{title}{Sub-projection-noise sensitivity in broadband atomic
  magnetometry}.
\newblock \emph{\bibinfo{journal}{Phys. Rev. Lett.}}
  \textbf{\bibinfo{volume}{104}}, \bibinfo{pages}{093602}
  (\bibinfo{year}{2010}).

\bibitem{koch07}
\bibinfo{author}{Koch, J.} \emph{et~al.}
\newblock \bibinfo{title}{Charge-insensitive qubit design derived from the
  Cooper pair box}.
\newblock \emph{\bibinfo{journal}{Phys. Rev. A}} \textbf{\bibinfo{volume}{76}},
  \bibinfo{pages}{042319} (\bibinfo{year}{2007}).

\bibitem{paik113D}
\bibinfo{author}{Paik, H.} \emph{et~al.}
\newblock \bibinfo{title}{Observation of high coherence in Josephson junction
  qubits measured in a three-dimensional circuit QED architecture}.
\newblock \emph{\bibinfo{journal}{Phys. Rev. Lett.}}
  \textbf{\bibinfo{volume}{107}}, \bibinfo{pages}{240501}
  (\bibinfo{year}{2011}).

\bibitem{cave82}
\bibinfo{author}{Caves, C.~M.}
\newblock \bibinfo{title}{Quantum limits on noise in linear amplifiers}.
\newblock \emph{\bibinfo{journal}{Phys. Rev. D}} \textbf{\bibinfo{volume}{26}},
  \bibinfo{pages}{1817--1839} (\bibinfo{year}{1982}).

\bibitem{cler10}
\bibinfo{author}{Clerk, A.~A.}, \bibinfo{author}{Devoret, M.~H.},
  \bibinfo{author}{Girvin, S.~M.}, \bibinfo{author}{Marquardt, F.} \&
  \bibinfo{author}{Schoelkopf, R.~J.}
\newblock \bibinfo{title}{Introduction to quantum noise, measurement, and
  amplification}.
\newblock \emph{\bibinfo{journal}{Rev. Mod. Phys.}}
  \textbf{\bibinfo{volume}{82}}, \bibinfo{pages}{1155--1208}
  (\bibinfo{year}{2010}).

\bibitem{koro11}
\bibinfo{author}{Korotkov, A.~N.}
\newblock \bibinfo{title}{Quantum Bayesian approach to circuit QED
  measurement}.
\newblock \emph{\bibinfo{journal}{arXiv:1111.4016}}  (\bibinfo{year}{2011}).

\bibitem{wisebook}
\bibinfo{author}{Wiseman, H.} \& \bibinfo{author}{Milburn, G.}
\newblock \emph{\bibinfo{title}{Quantum Measurement and Control}}
  (\bibinfo{publisher}{Cambridge University Press}, \bibinfo{year}{2010}).

\bibitem{schu05}
\bibinfo{author}{Schuster, D.~I.} \emph{et~al.}
\newblock \bibinfo{title}{AC Stark shift and dephasing of a superconducting
  qubit strongly coupled to a cavity field}.
\newblock \emph{\bibinfo{journal}{Phys. Rev. Lett.}}
  \textbf{\bibinfo{volume}{94}}, \bibinfo{pages}{123602}
  (\bibinfo{year}{2005}).

\bibitem{bois09}
\bibinfo{author}{Boissonneault, M.}, \bibinfo{author}{Gambetta, J.~M.} \&
  \bibinfo{author}{Blais, A.}
\newblock \bibinfo{title}{Dispersive regime of circuit QED: photon-dependent
  qubit dephasing and relaxation rates}.
\newblock \emph{\bibinfo{journal}{Phys. Rev. A}} \textbf{\bibinfo{volume}{79}},
  \bibinfo{pages}{013819} (\bibinfo{year}{2009}).

\bibitem{koro06}
\bibinfo{author}{Korotkov, A.~N.} \& \bibinfo{author}{Jordan, A.~N.}
\newblock \bibinfo{title}{Undoing a weak quantum measurement of a solid-state
  qubit}.
\newblock \emph{\bibinfo{journal}{Phys. Rev. Lett.}}
  \textbf{\bibinfo{volume}{97}}, \bibinfo{pages}{166805}
  (\bibinfo{year}{2006}).

\bibitem{katz06}
\bibinfo{author}{Katz, N.} \emph{et~al.}
\newblock \bibinfo{title}{Coherent state evolution in a superconducting qubit
  from partial-collapse measurement} \textbf{\bibinfo{volume}{312}},
  \bibinfo{pages}{1498--1500} (\bibinfo{year}{2006}).

\bibitem{katz08}
\bibinfo{author}{Katz, N.} \emph{et~al.}
\newblock \bibinfo{title}{Reversal of the weak measurement of a quantum state
  in a superconducting phase qubit}.
\newblock \emph{\bibinfo{journal}{Phys. Rev. Lett.}}
  \textbf{\bibinfo{volume}{101}}, \bibinfo{pages}{200401}
  (\bibinfo{year}{2008}).

\bibitem{kim11}
\bibinfo{author}{Kim, Y.-S.}, \bibinfo{author}{Lee, J.-C.},
  \bibinfo{author}{Kwon, O.} \& \bibinfo{author}{Kim, Y.-H.}
\newblock \bibinfo{title}{Protecting entanglement from decoherence using weak
  measurement and quantum measurement reversal}.
\newblock \emph{\bibinfo{journal}{Nature Physics}}
  \textbf{\bibinfo{volume}{8}}, \bibinfo{pages}{117--120}
  (\bibinfo{year}{2011}).

\bibitem{groe13}
\bibinfo{author}{Groen, J.~P.} \emph{et~al.}
\newblock \bibinfo{title}{Partial-measurement back-action and non-classical
  weak values in a superconducting circuit}.
\newblock \emph{\bibinfo{journal}{arXiv:1302.5147}}  (\bibinfo{year}{2013}).

\bibitem{wise12}
\bibinfo{author}{Wiseman, H.~M.} \& \bibinfo{author}{Gambetta, J.~M.}
\newblock \bibinfo{title}{Are dynamical quantum jumps detector dependent?}
\newblock \emph{\bibinfo{journal}{Phys. Rev. Lett.}}
  \textbf{\bibinfo{volume}{108}}, \bibinfo{pages}{220402}
  (\bibinfo{year}{2012}).

\end{thebibliography}


Correspondence and requests for materials should be addressed to katerm@berkeley.edu



\end{document}








\title{Supplementary information for  \\ "Observing single quantum trajectories of a superconducting qubit"}

\author{K. W. Murch}
\affiliation{Quantum Nanoelectronics Laboratory, Department of Physics, University of California, Berkeley CA 94720}
\author{S. J. Weber}
\affiliation{Quantum Nanoelectronics Laboratory, Department of Physics, University of California, Berkeley CA 94720}
\author{C.  Macklin}
\affiliation{Quantum Nanoelectronics Laboratory, Department of Physics, University of California, Berkeley CA 94720}
\author{I. Siddiqi}
\affiliation{Quantum Nanoelectronics Laboratory, Department of Physics, University of California, Berkeley CA 94720}


\date{\today}
\maketitle

 \begin{figure*}
 \begin{center}
\includegraphics[width=.8\textwidth]{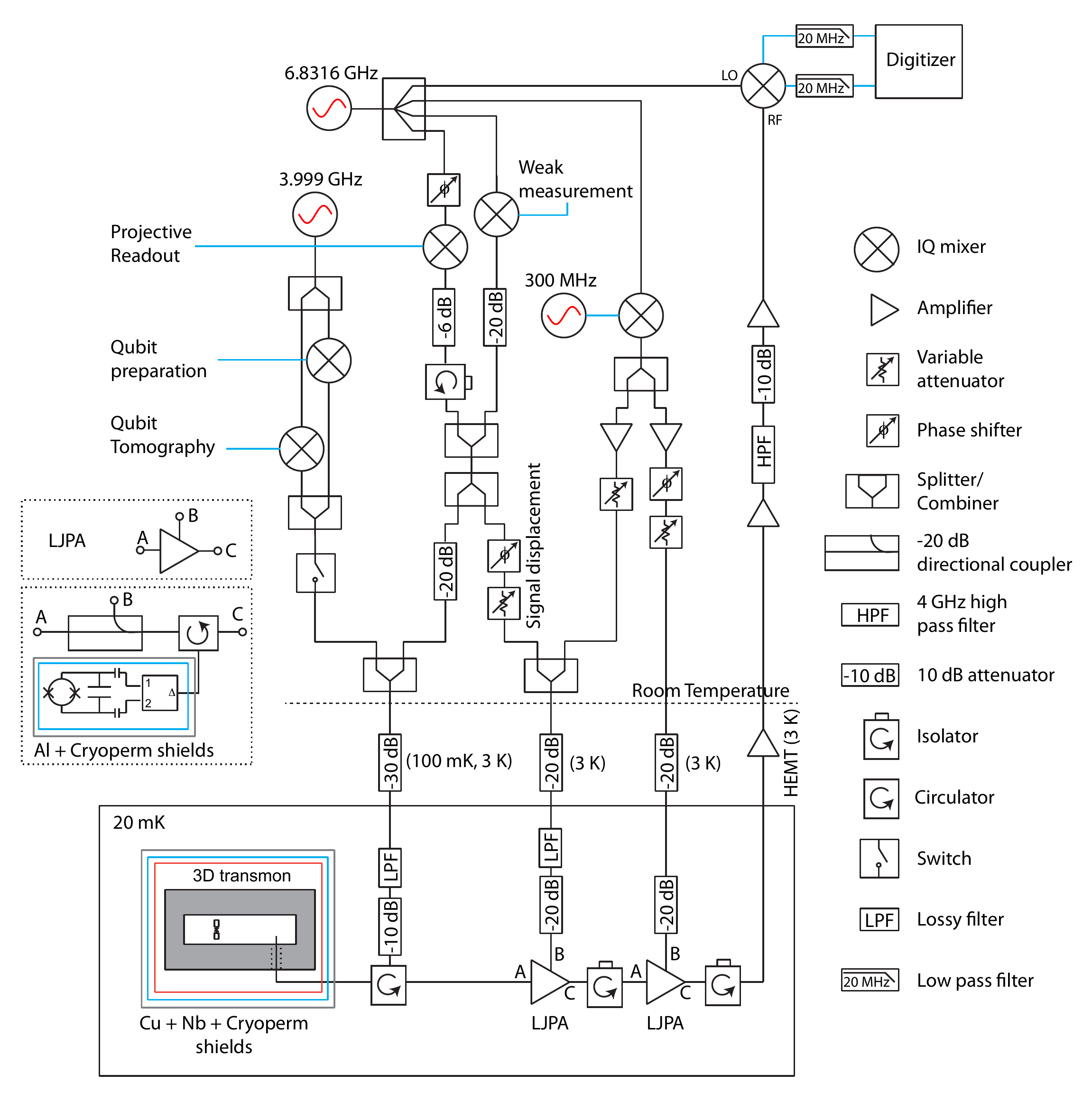}
\caption{\label{fig3} Experimental schematic.  The  weak measurement tone is always on, and has a variable amplitude and phase. The projective readout tone is pulsed, and adjusted to make a $Z$-measurement with $\mathcal{S} = 42$.  The amplitude and phase of the signal displacement tone are adjusted to displace the measurement signals back to the origin of the $X_1 X_2$ plane and allows the LJPA to perform in the linear regime.} 
\end{center}
\end{figure*}
\pagebreak

In principle, the process of measurement involves continuous entanglement of the measurement signal with the qubit state. As such, quadrature specific amplification of the measurement signal allows remote steering of the random motions of the qubit state induced by measurement--a notion known as ``EPR steering''.   Proof of EPR steering can be obtained by violation of the inequality\cite{wise12}
\begin{eqnarray}
S^{Z,\phi}\equiv\langle \z^Z\rangle^2 + \langle \x^\phi\rangle^2+\langle \y^\phi\rangle^2\le1.
\end{eqnarray}
Here, we define measurement-correlated ensemble averages, $\langle \x^\phi\rangle =\int \x^\phi  P(V_m) dV_m$,  where $P(V_m)$ is the probability distribution of measurement values.  We similarly define  $\langle \x^Z\rangle,\ \langle \z^Z\rangle$, and $\langle\y^\phi\rangle$.   As shown in Fig.\ \ref{figs3}, $\langle \z^Z\rangle$  increases with increasing $\nbar$ as a larger number of measurement outcomes result in projection of the qubit, thereby purifying the ensemble. Similarly, $\sqrt{\langle \x^\phi\rangle^2+\langle \y^\phi\rangle^2}$ is larger than $\langle \x^Z\rangle$ indicating that the $\phi$-measurement allows us to track phase shifts imparted by the fluctuating intracavity photon number. 

In our experiment we obtain $S^{Z,\phi}\sim 0.8$ due to measurement inefficiency and extra environmental dephasing.  We independently measured  $1.4$ dB of attenuation of our measurement signal between the cavity port and the parametric amplifier.  We separate the measurement efficiency $\eta = \eta_\textrm{col} \eta_\textrm{amp}$, where $\eta_\textrm{col} =0.72$ is the collection efficiency and $\eta_\textrm{amp} = 0.68$ is the amplifier quantum efficiency.  While $\eta_\textrm{amp}<1$ indicates that the amplifier does not operate with unity quantum efficiency, our observation that $\eta_\textrm{amp}>0.5$ indicates that the majority of the backaction of the measurement depends the quadrature of amplification. In this way, the amplification process  guides the stochastic evolution of the qubit state, but not to an extent that allows a demonstration of EPR steering. 

\begin{figure*}
 \begin{center}
\includegraphics[width=.5\textwidth]{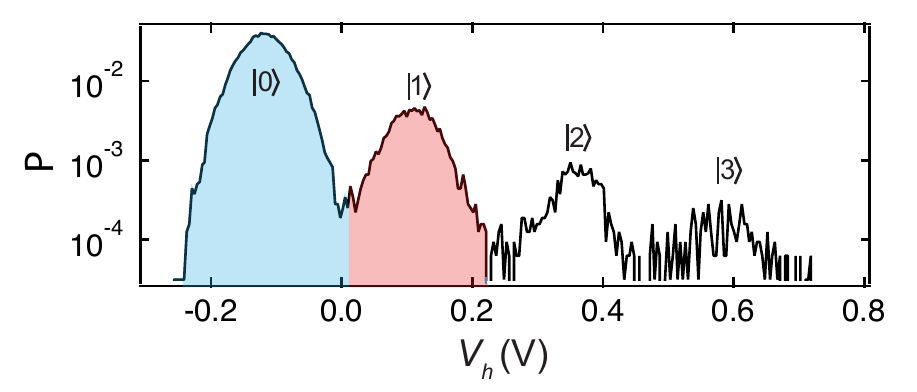}
\caption{\label{fig3} Single-shot, multi-state quantum non-demolition measurement. The graph shows the distribution of herald readout values $V_\mathrm{h}$. In the regime $|\chi|\ll\kappa$, several states of the transmon qubit are visible in the herald and tomography readout.  } 
\end{center}
\end{figure*}

 \begin{figure*}
 \begin{center}
\includegraphics[width=.3\textwidth]{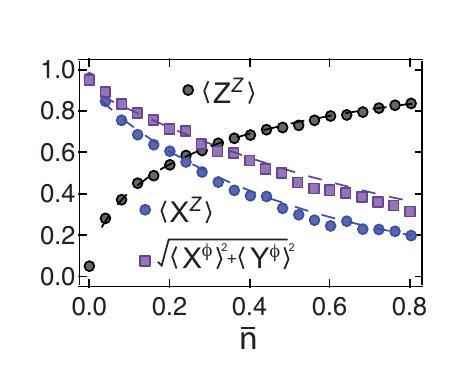}
\caption{\label{figs3} Measurement-correlated ensemble averages. } 
\end{center}
\end{figure*}
